\documentclass[twocolumn,showpacs,preprintnumbers,amsmath,amssymb]{revtex4}
\usepackage{graphicx}
\usepackage{dcolumn}
\usepackage{bm}
\begin{document}
\title {Asymptotic behavior of the
Kohn-Sham exchange potential at a metal surface}
\author {Zhixin Qian}
\affiliation{Department of Physics,
Peking University, Beijing 100871, China}
\date{\today}
\begin{abstract}
The asymptotic structure of the Kohn-Sham exchange potential $v_x ({\bf r})$
in the classically forbidden region of a metal surface 
is investigated, together with that of the Slater
exchange potential $V_x^S ({\bf r})$ and those of the approximate 
Krieger-Li-Iafrate $V_x^{KLI} ({\bf r})$ and Harbola-Sahni $W_x ({\bf r})$
exchange potentials. Particularly, the former is shown to have the
form of $v_{x} (z \to \infty) = -  \alpha_x / z$ with $\alpha_x$ a constant
dependent only of bulk electron density.
The same result in previous work is thus confirmed; in the meanwhile
controversy raised recently gets resolved.
The structure of the exchange hole
$\rho_x ({\bf r}, {\bf r}')$ is examined, and
the delocalization of it in the metal bulk when the electron
is at large distance from the metal surface 
is demonstrated with analytical expressions. The asymptotic structures
of $v_x ({\bf r})$,  $V_x^S ({\bf r})$, 
$V_x^{KLI} ({\bf r})$, and $W_x ({\bf r})$ at a slab metal surface are
also investigated. Particularly, $v_{x} (z \to \infty) = -  1/ z$ in the
slab case. The distinction in this respect between the semi-infinite
and the slab metal surfaces is elucidated. 
\end{abstract}

\maketitle

\section{Introduction}
The static (classical) charge potential vanishes exponentially
far outside a metal surface. The electronic 
structure in the classically forbidden
region as a consequence depends strongly on exchange and correlation (xc)
effects. Quantum mechanical many-body effects thus play a
major role in this region. 
In the seminal work of application of the Kohn-Sham (KS)
density functional theory (DFT) \cite{HK-KS}
to the metal surface problem, Lang and Kohn \cite{LK} assumed that the KS
local xc potential $v_{xc} (z)$ decays like the classical
image potential $V_{im}(z)=- 1/4z$, with $z$ being the distance
from the metal surface. (Similar assumption had been made earlier in 
Bardeen's pioneering study of the electronic properties of
the metal surface \cite{bardeen}.)
The KS local xc potential $v_{xc} ({\bf r})$ is equal to
the functional
derivative of the KS xc energy functional
$E_{xc} [\rho ]$ with respect to
the electron density $\rho ({\bf r})$: $v_{xc} ({\bf r}) 
= \delta E_{xc} [\rho]/ \delta \rho ({\bf r})$. 
It was found that $v_{xc} (z)$ calculated 
by Lang and Kohn in the local density approximation (LDA)  
to $E_{xc} [\rho ]$ turned out to have an exponential decay at 
large distance from the metal surface, recognized hence as one of 
shortcomings of the LDA \cite{LK}. 
The subject of the asymptote of $v_{xc} (z)$ at the metal surface has 
thereafter attracted 
long-standing research interest \cite{gunnarsson}. 
It was claimed in Ref. \cite{barth} and Ref. \cite{sham} 
that  $v_{xc} (z) \sim -1/(4z)$ which was
attributed to the correlation potential $v_{c} (z)$.
The exchange potential $v_{x} (z)$ was claimed to decay exponentially
in the former and as  $\sim
- 1/z^2$ in the latter, respectively. 

The exponential and the $\sim -1/z^2$ behavior of the exchange potential
$v_x (z)$ was questioned in Ref. \cite{HS1}, and it was numerically
demonstrated that at least the component of $v_x (z)$ arising purely from
the exchange hole, [which is in fact the Harbola-Sahni approximate 
exchange potential
$W_x (z)$ discussed later in this paper,] 
has an image-potential-like behavior though possibly not
the exact form of $V_{im} (z)$.
The result in Ref. \cite{HS1} was corroborated by Solomatin and Sahni \cite{SS1}
who, based on the integral equation for the optimized effective potential (OEP)
\cite{OEP} [also known as the OPM (optimized potential 
method) in \cite{SS1} and various literature],
analytically showed that 
\begin{eqnarray} \label{eq1}
v_{x} (z \to \infty) = -  \alpha_x \frac{1}{z},
\end{eqnarray}
with
\begin{eqnarray}  \label{alphax}
\alpha_x =  \frac{\beta^2 -1}{2 \beta^2} \biggl [1- \frac{
ln (\beta^2 -1)}{\pi (\beta^2 -1)^{1/2}} \biggr ] ,
\end{eqnarray}
where $\beta = \sqrt{W/\epsilon_F}$,
$W$ the surface-barrier height, and $\epsilon_F $ the bulk Fermi energy.
This result had been further confirmed in Ref. \cite{QS1}. The issue of
the asymptote of the full KS exchange-correlation potential $v_{xc} (z)$
was also addressed in Ref. \cite{QS1}.

Both the calculations in Refs. \cite{SS1} and \cite{QS1} for
$v_x (z)$ at large $z$ had been carried out exactly 
(with no approximation employed). The one in Ref. \cite{QS1} made the use of a
different method by solving the Dyson equation 
with the use of the exact exchange part
$\Sigma_x$ of the electron self-energy. The agreement between 
Ref. \cite{SS1} and Ref. \cite{QS1}  on the
result shown in Eq. (\ref{eq1}) strongly indicates its correctness.
Recently Horowitz et al. however
claimed \cite{horowitz1} that 
asymptotically $v_x (z)=-1/z$, but later \cite{horowitz2} 
that $v_x (z)$ has an 
asymptotic form of $v_x (z) \sim \ln z /z$. The result in Eq. (\ref{eq1}) was 
hence challenged. 
The method used in Refs. \cite{horowitz1,horowitz2} was  
the same OEP method used previously 
in Ref. \cite{SS1}. However there was a subtle but distinct difference
in the technique between the approach in Ref. \cite{SS1} and
that in Refs. \cite{horowitz1,horowitz2}. The calculation in Ref. \cite{SS1}
made the direct use of the integral equation for OEP, but those in 
Refs. \cite{horowitz1,horowitz2} the use instead 
of the OEP method formulated
in terms of differential equations for the so-called orbital-shifts [cf. 
Eq. (\ref{temp1}) below] \cite{OEP2,kummel,gorling}.  
There are certain advantages in this formulation
% in comparison to
%the integral equation formulation 
in that the quantities of the
orbital-shifts are comparatively amenable to analytical or numerical
study. This will also get 
illustrated in this work.  
Indeed, we attempt to resolve the controversy by
carrying out further investigation based precisely on this formulation of
the OEP method. To this end, we
have managed to establish, for limiting
large $z$, an identity between $v_x (z)$ and the planar-momentum 
averaged orbital-dependent exchange
potential $u_{xk} (z)$ at $k=k_F$, where $k$ is the perpendicular
component of the electron momentum, and $k_F$ the bulk Fermi momentum. 
The identity, reported in Eq. (\ref{ident}) [for the case of the slab metal 
surface see instead that in Eq. (\ref{ident0})], is one of the key results,
with the aid of which the following development can be made fairly smoothly.
The quantity $u_{xk_F} (z)$ is then shown to have
the asymptotic form of
$u_{xk_F} (z) = - \alpha_x  /z$.
In this manner, we confirm once again the result in Eq. (\ref{eq1}).
The controversy raised in Ref. \cite{horowitz2} is hence resolved.
It is shown in Sec. V that one component [termed
$V_x^{shift} (z)$] of $v_x (z)$ which was ignored in the study in 
Ref. \cite{horowitz2} actually also makes a leading-order contribution
to $v_x (z)$.
In addition, it is shown that $v_{x} (z=\infty) =0$ at the semi-infinite
metal surface, i.e., Eq.
(\ref{eq1}) is exact to the leading order. 
Since $v_{xc} (z= \infty) =0$ (which can always be made true), it follows
that $v_c (z= \infty) =0$.

The Slater exchange potential $V_x^S ({\bf r})$ has been regarded also
as an approximation to the KS exchange potential.
In Ref. \cite{SS1} it was shown that
$V_x^S (z)$ \cite{slater} has the following asymptotic
form at large distance from the metal surface:
\begin{eqnarray} \label{Vslater} 
V_x^S (z \to \infty) = -2 \alpha_x \frac{1}{ z}.
\end{eqnarray}
There exists no controversy in the literature over this result.
We give further verification of it.

The quite nontrivial asymptotic structure of $v_x (z)$ in Eq. (\ref{eq1})
was pointed out to arise from the delocalization of
the exchange hole (also known as the Fermi hole) 
$\rho_x ({\bf r}, {\bf r}')$ \cite{SB}. 
It was shown numerically that the exchange hole is
spread throughout the entire metal bulk when the electron
is in the classically forbidden 
region \cite{SB}. Expressions for the exchange
hole with the planar-positions averaged, valid for 
arbitrary electron positions,
are reported. 
Especially, the delocalization of 
the exchange hole at the semi-infinite metal surface
for limiting large electron positions
is illustrated with analytical expressions. This might be of help
to shed further light on the curious exchange effects in the classically
forbidden region of the metal surface.

Two of other approximate exchange potentials,
the Krieger-Li-Iafrate (KLI) $V_x^{KLI} ({\bf r})$
potential \cite{KLI1,KLI2} and the 
Harbola-Sahni (HS) $W_x ({\bf r})$ potential 
\cite{HS2}, are also surveyed.  They will be introduced in Sec. V.
$V_x^{KLI} ({\bf r})$ is well-known as an extensively 
employed substitute for the OEP.
It has the same
bulk limit as $v_x(z)$, a fact crucial for calculation practice for the
metal surface. 
$V_x^{KLI} (z)$ turns out to have an asymptotic form of $V_x^{KLI} (z) \sim
\ln z/z$ and hence
deviate from $v_x(z)$ in the classically forbidden region,
indicating that an improvement is required 
there. On the other side, the asymptotic behavior of
$W_x (z)$ has been given a full-fledged 
study in Ref. \cite{SS2}, and it was shown that
$W_x (z)$ deviates from $v_x (z)$ 
in both the metal bulk and the classically forbidden region, 
though relatively mildly in the
latter in the form of $W_x (z) \sim - \alpha_W /z$, where the constant
$\alpha_W $ is given in Eq. (\ref{alphaW}).

A great deal of work on the electronic structure at the metal
surface has been carried out on a jellium metal slab instead. 
The asymptotic behavior of $v_x (z)$ at large distance from the 
slab surface is also examined in this work, and it is 
shown that
\begin{eqnarray} \label{vxslab0}
v_x (z \to \infty) = - \frac{1}{z} . 
\end{eqnarray}
As mentioned previously, the same result was also reported in 
Ref. \cite{horowitz1}. In this sense it gets confirmed here.
It is necessary to remark that the study in Ref. \cite{horowitz1}
was performed on the slab metal surface, 
but clearly also aimed at obtaining knowledge
equally valuable for the semi-infinite metal surface. We, however, point out
that 
Eq. (\ref{vxslab0}) is valid only for the slab surface and can not
be naively extrapolated for the semi-infinite surface.
In fact, the slab surface can virtually be regarded as a finite
system and as a consequence Eq. (\ref{vxslab0}) can be obtained
by a direct multipole expansion of the Coulomb interaction.
Furthermore it is
shown that the result in 
Eq. (\ref{vxslab0}) is in fact subject to a possible difference
of nonzero constant, i.e., $v_{x} (z=\infty) \neq 0$ 
[see Eq. (\ref{vxslab2}) 
and the discussion below it], though it can be made to vanish
in an exchange-only calculation via a shift in
$v_x (z)$. 

In Sec. II we introduce the OEP method to the metal surface problem.
The properties of $v_x (z)$, $V_x^S (z)$, and 
the exchange energy density $\epsilon_x (z)$ are examined in Sec. III.
The properties of $\rho_x({\bf r}, {\bf r}')$ are examined in Sec. IV,
and those of $V_x^{KLI} (z)$ and $W_x (z)$ in Sec. V. 
The slab surface case is considered in Sec. VI.
In Appendix A we prove a mathematical statement of Eq. (\ref{ur=uz}) 
proposed in Sec. II. Appendix B contains the calculation for 
the asymptotic form of $u_{xk_F} (z)$, Appendix C that 
for $V_x^S (z)$, and Appendix D that for the planar-position averages 
of $\rho_x({\bf r}, {\bf r}')$. The paper is concluded in Sec. VII.

\section{Preliminaries}

In the jellium \cite{bardeen} and
structureless-pseudopotential \cite{perdew-shore}
models of a metal surface with a uniform positive background of charge
\begin{eqnarray}  \label{posit}
\rho_{+} (z)= \frac{k_F^3}{3 \pi^2} \theta (-z),
\end{eqnarray}
the KS orbitals are of the form
\begin{eqnarray}  \label{orbital}
\phi_{\bf k} ({\bf r}) = \sqrt{\frac{2}{A L}}
e^{i {\bf k}_{\parallel} \cdot {\bf x}_{\parallel}}
\phi_k (z) . 
\end{eqnarray}
In Eq. (\ref{posit}) (${\bf k}_{\parallel}, {\bf x}_{\parallel}$) 
are the planar components of the momentum and position,
and ($k, z$) the perpendicular components, i.e., ${\bf k} = $ 
${\bf k}_{\parallel}+$ $k {\bf e}_z$ and $ {\bf r} =$ 
${\bf x}_{\parallel}+$ $z {\bf e}_z$, 
where $ {\bf e}_z$ is the unit vector perpendicular to the metal surface. 
The magnitude of ${\bf k}$, on the other hand, will
be explicitly denoted with $|{\bf k}|$ ($=\sqrt{ k_{\parallel}^2 +k^2}$). 
$A$ and $L$ in Eq. (\ref{orbital}) denote 
the planar normalization area and the perpendicular normalization
length, respectively. 

\subsection{Preparatory materials}

The $\phi_k (z)$ obeys the differential equation
\begin{eqnarray} \label{diff}
\biggl [ - \frac{1}{2} \frac{\partial^2}{\partial z^2}
+ W + V(z) + v_{xc} (z)  \biggr ] \phi_k (z)
= \frac{1}{2} k^2 \phi_k (z) ,
\end{eqnarray}
where $V(z)$ is the static (classical) charge potential which vanishes
exponentially at large distance from the metal surface. 
$\phi_k (z)$ has the following
asymptotic forms,
\begin{subequations}
\begin{eqnarray} 
\phi_k (z) && \sim  
\sin [kz + \delta (k)] ~~~~~~for ~~~  z \to - \infty ,
\label{sine} \\ 
&& \sim P_k (z) e^{- \kappa z}  ~~~~~~~~~~ for  ~~~ z \to  \infty,
\end{eqnarray}
\end{subequations}
where   
$\delta (k) $ is the phase shift due to the metal surface,
$\kappa = \sqrt{2 W - k^2}$, and
$P_k (z)$ is a power function \cite{QS1}.
$\infty$ denotes the positive infinity.

The Dirac density matrix is defined as 
$\gamma_s ({\bf r}, {\bf r}') = 2 \sum_{i} \phi_i ({\bf r})
\phi_i^* ({\bf r}')$. At the metal surface, 
it has the following form:
\begin{eqnarray} \label{gamma5} 
\gamma_s ({\bf r}, {\bf r}') = \frac{1}{2 \pi^3} \int d {\bf k}
&& \theta (k_F - |{\bf k}|)   \nonumber \\
&& \phi_k^* (z) \phi_k (z') 
e^{i {\bf k}_{\parallel} \cdot
({\bf x}_\parallel' - {\bf x}_\parallel)} .
\end{eqnarray}
By the use of the identities:
\begin{eqnarray} \label{ident5} 
\int d {\hat {\bf q}} e^{-i 
{\bf q} \cdot ({\bf x}_{\parallel}
- {\bf x}_\parallel') }
= 2 \pi J_0 (q |{\bf x}_\parallel -{\bf x}'_\parallel|) ;
\end{eqnarray}
\begin{eqnarray}
\int_0^\lambda d k_{\parallel} k_{\parallel}
J_0 (k_{\parallel} |{\bf x}_{\parallel} - {\bf x}'_{\parallel}| )
= \lambda \frac{J_1 (\lambda |{\bf x}_{\parallel} -
{\bf x}'_{\parallel}|)}
{|{\bf x}_{\parallel} - {\bf x}'_{\parallel}|} ,
\end{eqnarray}
where $J_0$ and $J_1$ are the zeroth order and the first order
Bessel functions, respectively,
we can rewrite Eq. (\ref{gamma5}) as
\begin{eqnarray} \label{A-gamma}
\gamma_s ({\bf r}, {\bf r}') = \frac{2}{\pi^2}
\int_0^{k_F} dk \lambda \phi_k (z) \phi_k^* (z')
\frac{J_1 (\lambda |{\bf x}_{\parallel} - {\bf x}'_{\parallel}|)}
{|{\bf x}_{\parallel} - {\bf x}'_{\parallel}|} .
\end{eqnarray}
The expression for the electron density is
\begin{eqnarray} \label{den5}
\rho (z) = \frac{1}{\pi^2 } \int_0^{k_F} dk \lambda^2  |\phi_k (z)|^2 .
\end{eqnarray}
[The density is equal to the diagonal component of the density matrix, 
and Eq. (\ref{den5}) follows directly from Eq. (\ref{A-gamma}) 
with the aid of the fact that $J_1(x) \to 1/2x$ as $x \to 0$.]

Finally we formulate for the metal surface problem the exchange
hole which is defined as 
\begin{eqnarray} \label{exchangehole}
\rho_x({\bf r}, {\bf r}')=
-\frac{|\gamma_s ({\bf r}, {\bf r}')|^2}{2 \rho ({\bf r})} .
\end{eqnarray}
It follows from the substitution of Eq. (\ref{gamma5}) 
into Eq. (\ref{exchangehole}) that
\begin{eqnarray} \label{rhox5}
\rho_x ({\bf r}, {\bf r}') &&=  -  \frac{1}{8\pi^6 \rho(z)}
\int d {\bf k} \int d {\bf k}' \theta (k_F-|{\bf k}|) \nonumber \\ 
&& \theta (k_F - |{\bf k}'|) 
\Phi_{k,k'} (z, z')
e^{i ({\bf k}_{\parallel} - {\bf k}'_{\parallel}) \cdot
({\bf x}_\parallel' - {\bf x}_\parallel)} ,  \nonumber \\
\end{eqnarray}
where  
\begin{eqnarray}
\Phi_{k,k'} (z, z') = \phi_k^{*} (z) \phi_k (z')
\phi_{k'} (z) \phi_{k'}^{*} (z') .
\end{eqnarray}
Alternatively, the substitution of Eq. (\ref{A-gamma}) instead yields
\begin{eqnarray}  \label{rhox-add}
\rho_x ({\bf r}, {\bf r}') = - && \frac{2}{\pi^4 \rho (z) }
\int_0^{k_F} dk \int_0^{k_F} dk' 
\lambda \lambda ' \Phi_{k,k'} (z, z') \nonumber \\
&& \frac{J_1 (\lambda |{\bf x}_{\parallel} - {\bf x}'_{\parallel}|)
J_1 (\lambda ' |{\bf x}_{\parallel} - {\bf x}'_{\parallel}|}
{|{\bf x}_{\parallel} - {\bf x}'_{\parallel}|^2} ,   
\end{eqnarray}
where $\lambda= \sqrt{k_F^2 - k^2}$ and $\lambda '= \sqrt{k_F^2 - k'^2}$.

\subsection{The OEP method}

The exchange-only OEP \cite{OEP} has been proved to be equal to
the KS exchange potential $v_x ({\bf r})$ \cite{shaginyan}. 
We briefly outline
the method below, starting with an introduction to the orbital-dependent
exchange potentials \cite{OEP2,kummel}
\begin{eqnarray} \label{ux}
u_{x i} ({\bf r}) = \frac{1}{\phi_i ({\bf r})}
\frac{\delta E_x [\{\phi_j \}]}{\delta \phi_i^* ({\bf r})},
\end{eqnarray}
and the orbital shifts
\begin{eqnarray} \label{psi}
\psi_{ i} ({\bf r}) = \sum_{j \neq i}
\frac{\langle \phi_j | v_x -u_{xi} |\phi_i \rangle }
{\epsilon_i - \epsilon_j} \phi_{j} ({\bf r}).
\end{eqnarray}
Here $\phi_i ({\bf r})$ and $\epsilon_i$ are the KS
orbitals and the corresponding
eigenenergies, respectively,
and $E_x[\{ \phi_j \}]$ is the exchange energy functional
of the orbitals.
The orbital-dependent exchange potentials
$u_{xi} ({\bf r}) $ in Eq. (\ref{ux}) are of
the same form as the Hartree-Fock potentials
but constructed from the KS orbitals $\phi_i ({\bf r})$. Explicitly,
\begin{eqnarray} \label{HF-pot}
u_{x i} ({\bf r})= -\frac{1}{2 \phi_i ({\bf r})}
\int d {\bf r}' \frac{\gamma_s ({\bf r}, {\bf r}') \phi_i ({\bf r}')}
{|{\bf r} - {\bf r}' |} .
\end{eqnarray}
The central equation in the OEP method is
\begin{eqnarray} \label{oep0}
\sum_i^{occ.} \psi_i^* ({\bf r}) \phi_i ({\bf r}) + c.c.= 0,
\end{eqnarray}
where $c.c.$ denotes the complex conjugate of the previous term.
Equations (\ref{psi})
and (\ref{oep0}), together with
the KS equations for the orbitals [cf. Eq. (\ref{diff})] \cite{HK-KS},
build the self-consistent calculation scheme for the exchange-only OEP.

In self-consistent calculation for the OEP, 
evaluating $\psi_i ({\bf r})$
from Eq. (\ref{psi}) is highly impractical. 
Equation (\ref{psi}) therefore is usually
rewritten in the form of a differential
equation \cite{OEP2,kummel,gorling}. In our case,
\begin{eqnarray} \label{temp1}
\biggl [ - \frac{1}{2} \bigtriangledown^2
&+& W + V(z) + v_{xc} (z)  -\frac{1}{2}{\bf k}^2 \biggr ]
\psi_{\bf k} ({\bf r})   \nonumber \\
&=& [-v_x (z) +u_{x{\bf k}} ({\bf r}) +D_{\bf k} ] 
\phi_{\bf k} ({\bf r}),
\end{eqnarray}
where
\begin{eqnarray}  \label{D5} 
D_{\bf k} = \langle \phi_{\bf k} | v_x
- u_{x {\bf k}}| \phi_{\bf k} \rangle .
\end{eqnarray}
It is demonstrated in Appendix A that $u_{x{\bf k}} ({\bf r})$ 
is independent of ${\bf x}_\parallel$, i.e., it is
a function of $z$ only:
\begin{eqnarray} \label{ur=uz}
u_{x{\bf k}} ({\bf r})=u_{x{\bf k}} (z) .
\end{eqnarray}
The variables
${\bf x}_\parallel$ and $z$ accordingly can be separated
in Eq. (\ref{temp1}) and $\psi_{\bf k} ({\bf r})$ hence has the
form
\begin{eqnarray}  
\psi_{\bf k} ({\bf r}) = \sqrt{\frac{2}{A L}}
e^{i {\bf k}_{\parallel} \cdot {\bf x}_{\parallel}}
\psi_{\bf k} (z) .
\end{eqnarray}
Correspondingly, with the aid of
Eq. (\ref{orbital}), it follows that
\begin{eqnarray} \label{temp2}
\biggl [ - \frac{1}{2} \frac{\partial^2}{\partial z^2}
&+& W + V(z) + v_{xc} (z)  -\frac{1}{2}k^2 \biggr ]
\psi_{\bf k} (z)   \nonumber \\
&=& [-v_x (z) +u_{x{\bf k}} (z) +D_{\bf k} ] \phi_k (z).
\end{eqnarray}
We average over ${\bf k}_\parallel$ on both sides of
the preceding equation and obtain
\begin{eqnarray} \label{diff2}
\biggl [ - \frac{1}{2} \frac{\partial^2}{\partial z^2}
&+& W + V(z) + v_{xc} (z)  -\frac{1}{2}k^2 \biggr ]
\psi_k (z)   \nonumber \\
&=& [-v_x (z) +u_{xk} (z) +D_k ] \phi_k (z).
\end{eqnarray}
The planar-momentum averages $u_{x k}$ $(z)$
of the orbital-dependent
exchange potentials $u_{x {\bf k}}$ $({\bf r})$ [rewritten as
$u_{x{\bf k}}$ $(z)$ according to Eq. (\ref{ur=uz})],
\begin{eqnarray} \label{ux1}
u_{x k} (z) = \frac{1}{\pi \lambda^2 } \int_0^{\lambda}
d {\bf k}_{\parallel} u_{x {\bf k}} (z),
\end{eqnarray}
has been introduced into Eq. (\ref{diff2}) together with
the similar planar-momentum averages $\psi_{ k} (z)$ of
$\psi_{\bf k} (z)$, and $D_k$ of $D_{\bf k}$.
%have been introduced into Eq. (\ref{diff2}).

We write 
\begin{eqnarray}
D_{\bf k}^a = \langle \phi_{\bf k} | v_x | \phi_{\bf k} \rangle ; ~~~
~~~ 
D_{\bf k}^b = \langle \phi_{\bf k} | u_{x{\bf k}} | 
\phi_{\bf k} \rangle .
\end{eqnarray}
Accordingly, it follows from Eq. (\ref{D5}) that
\begin{eqnarray} \label{DDD}
D_k=D_k^a - D_k^b,
\end{eqnarray}
with
\begin{eqnarray}  \label{DDDD}
D_k^a= 2 \langle \phi_k |v_x |\phi_k \rangle / L ; ~~~ 
~~~
D_k^b= 2 \langle \phi_k |u_{xk} |\phi_k \rangle /L .
\end{eqnarray} 
For the semi-infinite metal surface, 
$D_k^a$ evidently takes 
the bulk value $v_x^{bulk} = -k_F/\pi$ of $v_x (z)$, (and is
accordingly independent of $k$.) On the other hand,
\begin{eqnarray} 
D_{\bf k}^b = - \frac{1}{2 \pi^2} \int d {\bf k}' \frac{1} {|{\bf k} - 
{\bf k}'|^2 } .
\end{eqnarray}
Explicitly \cite{bardeen,ashcroft},
\begin{eqnarray} 
D_{\bf k}^b = -\frac{ 2 k_F}{\pi}  F \biggl ( \frac{|{\bf k}|}{k_F} \biggr ) ,
\end{eqnarray}
with
\begin{eqnarray}
F (x) = \frac{1}{2} + \frac{1 - x^2}{4x} \ln |\frac{1+x}{1-x}| .
\end{eqnarray}
Solely for later reference, we also list
\begin{eqnarray} \label{Dk-vect}
D_{\bf k} = \frac{ k_F}{\pi} \biggl [
2 F \biggl ( \frac{|{\bf k}|}{k_F} \biggr ) - 1 \biggr ] ,
\end{eqnarray}
and note that $D_{\bf k} |_{|{\bf k}|= k_F}$ $ =0$.
Expression for $D_k^b$ will also be used in the 
later development:
\begin{eqnarray} 
D_k^b &=& -\frac{2k_F}{\pi^2 \lambda^2 } \int_0^\lambda
d {\bf k}_{\parallel}  F 
\biggl ( \frac{|{\bf k}|}{k_F} \biggr ) \nonumber \\
&=& - \frac{4k_F^3}{\pi \lambda^2 } \int_{k/k_F}^1
d x x F(x) ;
\end{eqnarray}
explicitly:
\begin{eqnarray} \label{Dkb}
D_k^b = \frac{1}{3 \pi} \biggl [&& - 2k_F 
+ \frac{(k_F +k) (2k_F -k)}{k_F- k}  \ln \frac{k_F +k}{2k_F}  \nonumber \\
&&+ \frac{(k_F -k) (2k_F+k)}{k_F +k} 
\ln \frac{k_F -k}{2k_F} \biggr ].
\end{eqnarray}

\subsection{Identity $v_x (z) =u_{x k_F} (z)$ for $z \to \infty$}

We consider Eq. (\ref{diff2}) for limiting large $z$. 
On the rhs of it the term $u_{xk} (z) \phi_k (z)$ which is due to
the orbital-dependent exchange potential has the asymptotic
form of $f_k (z) \phi_{k_F} (z)$, with $f_k (z)$ a power function.
Its existence implies that $\psi_k (z)$ must 
have the
analogous asymptotic form:
\begin{eqnarray} \label{psi2}
\psi_k (z)=g_k(z) \phi_{k_F} (z),
\end{eqnarray}
where $g_k (z)$ is also a power function obeying
\begin{eqnarray} \label{gk0}
\frac{1}{2} \lambda^2 && g_k(z) - 
\phi_{k_F}^{-1} (z)  \phi_{k_F}' (z) g_k '(z)
-\frac{1}{2}  g_k'' (z) \nonumber \\
&&= \phi_{k_F}^{-1} (z) [-v_x (z) +u_{xk} (z) +D_k ] \phi_k (z).
\end{eqnarray}
The primes denote the derivatives with respect to $z$.
Equation (\ref{gk0}) is valid for all $k$. It turns out
that its special case at $k=k_F$ solely 
is sufficient to determine the asymptotic
structure of $v_x (z)$.
Indeed,
\begin{eqnarray} \label{gk}
\phi_{k_F}^{-1} (z)  \phi_{k_F}' (z) g_{k_F} '(z)
&+& \frac{1}{2}  g_{k_F}'' (z)   \nonumber \\
&=& v_x (z) - u_{xk_F} (z) - D_{k_F} .
\end{eqnarray} 
By using the fact that $g_{k_F} (z) \to 0$, [with the
understanding that the homogeneous
solution of $\psi_k (z)$ in Eq. (\ref{diff2}) is excluded,]
one has $g_{k_F} '(z)$, $g_{k_F}'' (z)$ $\sim o(1/z)$ for large
$z$. On the other hand, it will be shown later in Eq. (\ref{uxkf})
that $u_{x k_F} (z)$ $\sim O(1/z)$. 
One hence obtains, from Eq. (\ref{gk}), for limiting large $z$
\begin{eqnarray}  \label{ident0}
v_x (z) =u_{x k_F} (z) + D_{k_F} .
\end{eqnarray}
Since $D_{k}^a = - k_F/\pi$, and $D_{k_F}^b = - k_F/\pi$ 
according to Eq. (\ref{Dkb}), one has correspondingly, 
from Eq. (\ref{DDD}), $D_{k_F}=0 $ for the semi-infinite metal surface. 
This leads to one of
the central identities:
\begin{eqnarray} \label{ident}
v_x (z) =u_{x k_F} (z) , ~~~~~for ~~ z \to \infty .
\end{eqnarray}
We remarked that $D_{k_F} \neq 0$ for the slab 
metal surface, and we then have Eq. (\ref{ident0}) only.

Equation (\ref{ident}) suggests that knowledge about the 
quantity $u_{x k_F} (z)$ could be useful
in carrying through the reminder of our task. In this regard we 
include here some expressions for it.
In general, for $k \to k_F$ and any well-behaved
function $f({\bf k})$ $=f({\bf k}_\parallel, k)$
[$u_{x {\bf k}} (z)$ in the present case],
\begin{eqnarray} \label{f}
\int_0^\lambda
d {\bf k}_{\parallel} f ({\bf k}) = \pi \lambda^2
f ({\bf k}_\parallel =0, k=k_F).
\end{eqnarray}
Hence from Eq. (\ref{ux1}), it follows that
\begin{eqnarray}  \label{uuident}
u_{x k_F} (z) = u_{x {\bf k}} (z)|_{{\bf k}_\parallel =0, k=k_F}.
\end{eqnarray}
Accordingly from Eqs. (\ref{HF-pot}) and (\ref{ur=uz}) one has 
\begin{eqnarray} \label{A-ux}
u_{x k_F} (z)= -\frac{1}{2 \phi_{k_F} (z)}
\int d {\bf r}' \frac{\gamma_s ({\bf r}, {\bf r}') \phi_{k_F} (z')}
{|{\bf r} - {\bf r}' |}. 
\end{eqnarray}

\section{Asymptotic structures of $v_x (z)$ and $V_x^S (z)$}

Quantities similar to $u_{x k_F} (z)$ have been calculated
in Refs. \cite{HS1,SS1,QS1}. We apply a different
approach to perform the calculation, which can serve also
as a kind of verification.
To this end, we substitute Eq. (\ref{A-gamma})
into Eq. (\ref{A-ux}).
It follows that
\begin{eqnarray} \label{A30}
u_{x k_F} (z)= &&-\frac{2}{\pi \phi_{k_F} (z)}
\int_0^{k_F} dk \lambda \phi_k (z) \nonumber \\
&&\int_{-\infty}^\infty d z' \phi_k^* (z')
\phi_{k_F} (z') \nonumber \\
&&\int_0^\infty dr_{\parallel}' \frac{1}
{\sqrt{(z -z')^2 +r_{\parallel}'^2}} J_1 (\lambda r_{\parallel}') .
\end{eqnarray}
Alternatively Eq. (\ref{A30}) can also be obtained from
the substitution of Eq. (\ref{ux5}) into Eq. (\ref{uuident}).
The aid of the fact that $J_0 (0) =1$ is needed then.

The asymptotic form of $u_{xk_F} (z)$ is examined in Appendix B
with the following result,
\begin{eqnarray}  \label{uxkf}
u_{x k_F} (z \to \infty)  = -\alpha_x \frac{1}{ z},
\end{eqnarray}
which, together with Eq. (\ref{ident}), 
leads to Eq. (\ref{eq1}), fulfilling the main object of this work.
(The controversy raised in Ref. \cite{horowitz2} will get further 
elucidated in Sec. V.)
Equation (\ref{uxkf}) which is exact
at limiting large $z$, together with Eq. (\ref{ident}), leads also to the 
conclusion that $v_x (\infty )$ $=0$  at the
semi-infinite metal surface.
In other words, Eq. (\ref{eq1}) is exact to
the leading order. We assume that $v_{xc} (\infty) =0 $ 
at the metal surface. Accordingly we also have $v_c (\infty )$ $=0$.
Further insight into these facts will be provided in Sec. VI 
in connection with those for the slab case.

The Slater exchange potential $V_x^S ({\bf r})$ \cite{slater} 
is defined as
\begin{eqnarray} \label{slater-def}
V_x^S ({\bf r})= \int d {\bf r}' \rho_x({\bf r}, {\bf r}') 
\frac{1}{| {\bf r} - {\bf r}'|}.
\end{eqnarray}
For our metal surface problem we substitute 
Eq. (\ref{rhox-add}) for $\rho_x$ $({\bf r}, {\bf r}')$ into 
Eq. (\ref{slater-def}). It follows that
\begin{eqnarray}   \label{vslat0}
V_x^S (z) = &&- \frac{4}{\pi^3 \rho (z)} \int_0^{k_F} dk \phi_k (z)
\int_0^{k_F} dk' \phi_{k'}^* (z) \lambda \lambda'  \nonumber \\
&& \int_{-\infty}^{\infty} d z'   \phi_k^* (z')
\phi_{k'} (z')   \nonumber \\
&&\int_0^\infty dr_{\parallel}' \frac{1}
{ r_{\parallel}'
\sqrt{(z -z')^2 +r_{\parallel}'^2}} J_1 (\lambda r_{\parallel}')
J_1 (\lambda' r_{\parallel}') .   \nonumber \\
\end{eqnarray}
The asymptotic structure of 
$V_x^S (z)$ at large distance from the metal surface is calculated
in Appendix C and the final result
is shown in Eq. (\ref{Vslater}) \cite{SS1}. 
Correspondingly, the exchange energy density per unit volume
$\epsilon_x ({\bf r}) = \frac{1}{2} \rho ({\bf r}) V_x^S ({\bf r})$
has the asymptotic form \cite{QS1}:
\begin{eqnarray}
\epsilon_x (z \to \infty) =  -\alpha_x \frac{\rho(z)}{z} .
\end{eqnarray}
We note that 
\begin{eqnarray}  \label{density99}
\rho(z \to \infty) = \frac{k_F}{2 \pi^2 c^2} \frac{1}{z^2}
|\phi_{k_F} (z)|^2 . 
\end{eqnarray}
%Equation (\ref{Vslater})
%is further confirmed in Appendix C.
 
It is well known that in a finite system $v_x ({\bf r})$ has a long-range
form of $v_x ({\bf r} \to \infty)$ $ = -1/r$ \cite{barth,sham,LPS,QS3}.
(For simplicity, only a spherically symmetric
system is discussed here. For recent progress made on
this subject, see Refs. \cite{gorling,kummel} and the discussions later
in Sec. VI.)
The Slater exchange potential has exactly the same
long-range form of $V_x^S ({\bf r} \to \infty)$ $ = -1/r$. The identical 
long-range form of $v_x ({\bf r})$ and $V_x^S ({\bf r})$ 
results from the fact that in the finite system
the exchange hole is well localized near the 
system. The asymptotic structure 
of $v_x ({\bf r})$ is essentially determined 
by that of the orbital-dependent
exchange potential $u_{xm}({\bf r})$ 
of the highest occupied orbital (denoted by
$m$) \cite{barth,sham,LPS,QS3}. The asymptotic structure of 
both $u_{xm}({\bf r})$ and $V_x^S ({\bf r})$ in turn
can be attributed to the (orbital-dependent) exchange hole in terms 
of the leading order contribution of the multipole expansion
of the Coulomb interaction. For the semi-infinite
metal surface this is no more
true since the exchange hole is delocalized and spread 
throughout the entire bulk region
\cite{SB}, and as a consequence the multipole expansion argument does not 
apply anymore. This explains the difference in the asymptotic
structures of $v_x (z)$ and $V_x^S (z)$ 
at large $z$ from the metal surface, as shown in Eq. (\ref{eq1}) and Eq.
(\ref{Vslater}), respectively. The delocalization of the exchange hole 
at large electron positions will be examined in the next section.

\section{Delocalization of $\rho_x({\bf r}, {\bf r}')$}

We take planar-position average
of $\rho_x({\bf r}, {\bf r}')$:
\begin{eqnarray} \label{rhox}
\rho_x (z, z') = \frac{1}{A} \int d {\bf x}_{\parallel}
\int d {\bf x}'_{\parallel} \rho_x({\bf r}, {\bf r}').
\end{eqnarray}
The information about $\rho_x (z, z')$ clearly is sufficient 
to serve the present purpose.
By the substitution of Eq. (\ref{rhox5}) into Eq. (\ref{rhox}),
$\rho_x (z, z')$ can be shown as 
\begin{eqnarray} \label{rhox1}
\rho_x (z, z')= && -\frac{2}{\pi^3 \rho(z)}
\int_0^{k_F} dk \lambda^2 \nonumber \\
&& \int_0^{k} d k'
[\Phi_{k,k'} (z, z') + \Phi_{k,k'} (z', z) ] .
\end{eqnarray}
Equation (\ref{rhox1}) is valid
for all electron positions $z$. This expression was first reported in 
Ref. \cite{SB} and the reader is referred to that work for detailed 
derivations. It is not difficult to see that 
the two terms in Eq. (\ref{rhox1}) make the same contribution, and hence
\begin{eqnarray}   \label{twoterm}
\rho_x (z, z')=  -\frac{4}{\pi^3 \rho(z)}
&& \int_0^{k_F} dk \lambda^2 \phi_k^* (z) \phi_k (z')   \nonumber \\
&& \int_0^{k} d k' \phi_{k'} (z) \phi_{k'}^* (z') .
\end{eqnarray}

We are interested especially in the asymptotic structure 
of $\rho_x (z, z')$. It is demonstrated in Appendix D that
\begin{eqnarray} \label{rhox2}
\rho_x (z \to \infty, z')=-\frac{4}{\pi c }\frac{1}{z}
|\phi_{k_F} (z')|^2 .
\end{eqnarray}
Equation (\ref{rhox2}) displays remarkably the 
delocalization of the exchange hole in the metal bulk. [The reader is
referred to Eq. (\ref{sine}) for the behavior of
$\phi_{k_F} (z)$ in the deep bulk region.]
In the meanwhile, the amplitude of $\rho_x (z, z')$ decays
as $\sim 1/z$ with the electron position $z$.

Both planar positions ${\bf x}_{\parallel}$ and 
${\bf x}'_{\parallel}$ of the electron and 
the hole have been averaged, respectively, in Eq. (\ref{rhox}). 
But evidently $\rho_x({\bf r}, {\bf r}')$ is a function of
${\bf x}_{\parallel}$$-{\bf x}'_{\parallel}$ only, rather than of
${\bf x}_{\parallel}$ and ${\bf x}'_{\parallel}$ separately.
Another type of average over the planar positions therefore might 
be equally capable of featuring the exchange hole at the metal 
surface which is defined in the following manner \cite{pitarke},
\begin{eqnarray} \label{bx}
b_x (z, z') = \int_0^\infty d 
|{\bf x}_{\parallel} - {\bf x}'_{\parallel}| \rho_x({\bf r}, {\bf r}').
\end{eqnarray}
We substitute Eq. (\ref{rhox5}) into Eq. (\ref{bx}) and write $b_x (z, z')$ 
as
\begin{eqnarray}
b_x (z, z')  &&= -\frac{1}{8 \pi^6 \rho (z)} \int_0^\infty
d |{\bf x}_\parallel -{\bf x}'_\parallel|  \nonumber \\
&& \int d{\bf k}
\int d{\bf k}' \theta (k_F - |{\bf k}|)
\theta (k_F - |{\bf k}'|) \nonumber \\
&&\Phi_{k,k'} (z, z') e^{-i
({\bf k}_{\parallel}
- {\bf k}_\parallel') \cdot ({\bf x}_{\parallel}
- {\bf x}_\parallel') } ,
\end{eqnarray}
By introducing the transforms: ${\bf q} = {\bf k}_{\parallel}
- {\bf k}'_{\parallel}$ and ${\bf K} = ( {\bf k}_{\parallel}
+ {\bf k}'_{\parallel})/2$, one may rewrite $b_x (z, z')$ as
\begin{eqnarray} \label{te9}
 b_x (z, z')  && = -\frac{1}{8 \pi^6 \rho (z)} \int_{-k_F}
^{k_F} dk \int_{-k_F}
^{k_F} dk' \Phi_{k,k'} (z, z') \nonumber \\
&& \int d {\bf q} F(q) \int_0^\infty
d |{\bf x}_\parallel -{\bf x}'_\parallel|
e^{-i
{\bf q} \cdot ({\bf x}_{\parallel}
- {\bf x}_\parallel') } ,
\end{eqnarray}
where the function $F(q)$
is defined as
\begin{eqnarray}  \label{F(q)0}
F(q) = \int d {\bf K} \theta (\lambda - |{\bf K}
+ {\bf q}/2 |) \theta (\lambda' - |{\bf K}
- {\bf q}/2 |) .
\end{eqnarray}
By using Eq. (\ref{ident5}) and further the following identity:
\begin{eqnarray}
\int_0^\infty d |{\bf x}_\parallel -{\bf x}'_\parallel|
J_0 (q |{\bf x}_\parallel -{\bf x}'_\parallel|)
= \frac{1}{q} ,
\end{eqnarray}
Eq. (\ref{te9}) may be further rewritten as
\begin{eqnarray} \label{b4}
b_x (z, z')=&& -\frac{1}{\pi^5 \rho(z)} \int_0^{k_F} dk
\nonumber \\
&& \int_0^{k_F} dk'
\Phi_{k,k'} (z, z')
\int_0^\infty dq F(q).
\end{eqnarray}
%We have also managed to calculate $b_x (z, z')$, and it is derived

Derivation based on Eq. (\ref{b4}), detailed in Appendix D, 
yields the following result for $b_x (z, z')$: 
\begin{eqnarray} \label{bx1}
b_x (z, z') = -\frac{16}{3 \pi^5 \rho (z) } 
[G(z, z') + G(z', z)],
\end{eqnarray}
where
\begin{eqnarray}  \label{G}
G(z, z') && =  \int_0^{k_F} dk \int_k^{k_F} dk'
\Phi_{k,k'} (z, z')   \nonumber \\
&&\lambda \biggl [
 {\bf K} \biggl (\frac{\lambda'}{\lambda} \biggr )
(\lambda'^2 - \lambda^2)    
+{\bf E} \biggl (\frac{\lambda'}{\lambda} \biggr )
(\lambda^2 + \lambda'^2) \biggr ] .  \nonumber \\
\end{eqnarray}
${\bf K}$ and ${\bf E}$ are the complete elliptic integrals of
the first and the second kinds, respectively \cite{gradshteyn}:
\begin{eqnarray}
{\bf K} (t) = \int_0^1 \sqrt{\frac{1}{(1-t^2x^2)(1-x^2)}} dx,
\end{eqnarray}
\begin{eqnarray}
{\bf E} (t) = \int_0^1 \sqrt{\frac{1-t^2x^2}{1-x^2}} dx .
\end{eqnarray}
We note that so far all results for $b_x (z, z')$
hold in general for arbitrary
position $z$. 
Particularly, Eq. (\ref{bx1}) evolves into the following asymptotic form
for limiting large electron position $z$, (see again Appendix D for 
the demonstration)
\begin{eqnarray} \label{bx2}
b_x (z \to \infty, z')=-\frac{\sqrt{ k_F } \gamma}{c^{3/2}}
 \frac{1}{z^{3/2}} |\phi_{k_F}(z')|^2,
\end{eqnarray}
where
\begin{eqnarray} \label{gamma}
\gamma = \frac{20 \sqrt{2}}{ \pi^{5/2}}  && \int_0^1 dy
( y +1)^{-7/2}   \nonumber \\
&& \times [{\bf K}(\sqrt{y}) (y-1) + {\bf E} (\sqrt{y}) (y +1)].
\end{eqnarray}
Equation (\ref{bx2})
displays similar delocalization nature of the exchange hole. The amplitude
of $b_x (z, z')$ decays as $\sim 1/z^{3/2}$, a little faster than that 
of $\rho_x (z, z')$ shown in Eq. (\ref{rhox2}).

\section{Asymptotic structure of approximate exchange potentials}

The Slater potential $V_x^S ({\bf r})$ was proposed as an approximation
to the orbital-dependent Hartree-Fock exchange potentials \cite{slater}. 
Upon the arrival
of the KS-DFT, $V_x^S ({\bf r})$ was regarded 
naturally also as an approximation
to the KS exchange potential $v_x ({\bf r})$. The large-distance structure
of it at the metal surface is shown in Eq. (\ref{Vslater}) and
discussed in Sec. III.
In this section we survey those of two of other approximate
exchange potentials, the KLI potential
$V_x^{KLI} ({\bf r}) $
and the HS potential $W_x ({\bf r})$. 
The former reads \cite{KLI1,KLI2}
\begin{eqnarray} \label{kli}
V_x^{KLI} ({\bf r})= V_x^S ({\bf r}) + V_x ^\bigtriangleup ({\bf r}) ,
\end{eqnarray}
where
\begin{eqnarray} \label{vdelta}
V_x^\bigtriangleup ({\bf r})= \frac{2}{\rho ({\bf r})} 
\sum_i^{occ.} |\phi_i ({\bf r})|^2
\langle \phi_i | v_x -u_{xi} | \phi_i \rangle .
\end{eqnarray}
In passing we mention that \cite{KLI2}
\begin{eqnarray} \label{vx-vkli}
v_x ({\bf r}) = V_x^{KLI} ({\bf r}) + 
V_x^{shift} ({\bf r}) ,
\end{eqnarray}
and the component $V_x^{shift} ({\bf r})$, defined as
\begin{eqnarray}
V_x^{shift} ({\bf r})= \frac{1}{\rho ({\bf r})}
\sum_i^{occ.} && [ \phi_i^* ({\bf r}) 
\bigtriangledown^2 \psi_i ({\bf r})  \nonumber \\
&& - \psi_i ({\bf r}) \bigtriangledown^2 
\phi_i^* ({\bf r}) ] ,
\end{eqnarray}
is sacrificed for the purpose of less calculation labor. [For convenience,
the same symbols $V_x ^\bigtriangleup ({\bf r})$ and
$V_x^{shift} ({\bf r})$ in Ref. \cite{horowitz2} are adopted here.]

We first make a digression to comment on some bulk properties of
$\psi_k (z)$ and $V_x^{KLI} (z)$. The expression for $V_x^{shift}(z)$ 
at the metal surface is
\begin{eqnarray} \label{bulk-sh}
V_x^{shift}(z) && =  \frac{1}{2 \pi^2 \rho(z)}
\int_0^{k_F} dk \lambda^2 \nonumber \\
&& \biggl [ 
\phi_k^* (z) \frac{\partial^2}{\partial z^2} \psi_k (z)
-\psi_k (z) \frac{\partial^2}{\partial z^2} \phi_k (z) \biggr ] .
\end{eqnarray}
On the other hand, in view of Eq. (\ref{DDDD}), one has
\begin{eqnarray}
v_x (z = -\infty) = D_k^a = -\frac{k_F}{\pi} ,
\end{eqnarray}
and 
\begin{eqnarray}
u_{xk} (z = -\infty) = D_k^b .
\end{eqnarray}
It follows that the rhs of Eq. (\ref{diff2}) vanishes. 
With the understanding that the homogeneous
solution of $\psi_k (z)$ in Eq. (\ref{diff2}) is excluded, one thus has
\begin{eqnarray}
\psi_k (z) =0 ~~~~~~~~~~~~~~~~~~ for ~~ z = -\infty.
\end{eqnarray}
Therefore $V_x^{shift} (z)$, according to Eq. (\ref{bulk-sh}), 
vanishes in the metal bulk:
\begin{eqnarray}
V_x^{shift} (z) = 0 ~~~~~~~~~~~~~ for ~~ z = -\infty ,
\end{eqnarray} 
and consequently $V_x^{KLI} (z)$, according to Eq. (\ref{vx-vkli}), 
has the merit of possessing
the same bulk limit as $v_x (z)$:
\begin{eqnarray} 
V_x^{KLI} (z) = - \frac{k_F}{ \pi} ~~~~~~~~ for ~~ z = -\infty .
\end{eqnarray}

We next return to the issue of the asymptotic behavior
of $V_x^{KLI} (z)$
in the classically forbidden region.
$V_x^{KLI} (z)$ is shown below to deviate
strongly from $v_x (z)$ at large distance from the metal surface. 
To this end, we first write the expression for
$V_x^\bigtriangleup (z)$:
\begin{eqnarray}
V_x^\bigtriangleup (z) = \frac{1}{\pi^2 \rho (z)}
\int_0^{k_F} dk \lambda^2 |\phi_k (z)|^2 D_k .
\end{eqnarray}
For limiting large $z$, $V_x^\bigtriangleup (z)$
turns out to have the form:
\begin{eqnarray} \label{vdelta2}
V_x^\bigtriangleup (z \to \infty) 
= \frac{\sqrt{\beta^2 -1}}{2 \pi} \frac{1}{z}
\biggl [&& \ln \biggl (\frac{k_Fz}{\sqrt{\beta^2 -1}} 
\biggr )  \nonumber \\
&& + C + 2 \ln 2 -1 \biggr ] ,  
\end{eqnarray}
with $C=0.577215$ the Euler constant. In obtaining
Eq. (\ref{vdelta2}), the following fact 
\begin{eqnarray} \label{te4}
D_k = \frac{1}{4 \pi} (k_F -k ) \biggl [ 
1 - 2 \ln \frac{k_F -k }{2 k_F} \biggr ]
+ o (k_F - k),
\end{eqnarray}
for small $k_F - k$, [which follows from Eq. (\ref{DDD})],
has been employed. Equation (\ref{vdelta2}) had also
been reported in Ref. \cite{horowitz2}. 
Finally, Eq. (\ref{kli}), together with Eqs. (\ref{Vslater}) and 
(\ref{vdelta2}), yields 
\begin{eqnarray}
V_x^{KLI} (z \to \infty) = && \frac{\sqrt{\beta^2 - 1}}{2 \pi} \frac{1}{z}
[\ln \biggl ( \frac{k_Fz}{\sqrt{\beta^2 -1}} \biggr ) \nonumber \\
&& + C + 2 \ln 2 -1 \biggr ]  
- 2 \alpha_x \frac{1}{z} ,
\end{eqnarray}
which has the leading form of $O(\ln z/z)$, 
and hence deviates from $v_x(z)$. 

The HS potential $W_x ({\bf r})$ \cite{HS2} 
is defined as the work 
done in the Pauli field ${\bf \cal E}_x ({\bf r})$ [$W_x ({\bf r})$
hence also known as the Pauli potential in the quantal density
functional theory (Q-DFT)] \cite{QDFT}, 
\begin{eqnarray} \label{wx0}
W_x ({\bf r} )= - \int_\infty^{\bf r} {\bf \cal E}_x ({\bf r}') \cdot
d {\bf l}',
\end{eqnarray}
and
\begin{eqnarray} \label{ex-curl}
{\bf \cal E}_x ({\bf r}) = - \int \rho _x 
({\bf r}, {\bf r}') \bigtriangledown \frac{1}{ | {\bf r}- {\bf r}'|} 
d {\bf r}' .
\end{eqnarray} 
The asymptotic structure of $W_x (z)$ has been extensively investigated 
in Ref. \cite{SS2}, and one of the main results is
\begin{eqnarray}  \label{Wx}
W_x (z) = - \alpha_W \frac{1}{z} ,
\end{eqnarray}
where
\begin{eqnarray}   \label{alphaW}
\alpha_W  &&=  \frac{\beta^2 - 1}{\beta^2}
\biggl [ \frac{\beta^2 -2 }{\beta^2 } 
+ \frac{2}{\pi \sqrt{\beta^2 - 1}}   \nonumber \\
&& \times 
\biggl ( 1 - \frac{(\beta^2 - 1) \ln (\beta^2 - 1 )}{\beta^2} \biggr )
\biggr ] .
\end{eqnarray}
It might be helpful to mention the relation
between $W_x ({\bf r})$ and $v_x ({\bf r})$ \cite{LM-QS}:
\begin{eqnarray}  \label{vx-Wx}
 v_x ({\bf r}) = W_x ({\bf r}) 
- W_{t_C}^{(1)} ({\bf r}),
\end{eqnarray}
where $W_{t_C}^{(1)} ({\bf r})$ is the lowest-order correlation-kinetic 
component of $ v_x ({\bf r})$. $W_x ({\bf r})$ has been
well recognized as the component of $v_x ({\bf r})$ which arises 
purely from the exchange hole $\rho _x
({\bf r}, {\bf r}')$ \cite{HS2,LM-QS}. 
Clearly $W_{t_C}^{(1)} (z)$ must also be long-ranged 
at the metal surface. This is in sharp contrast to its behavior
in finite systems such as atoms 
and molecules in which $W_{t_C}^{(1)} ({\bf r})$
decays in a rather short-ranged form of $\sim 1/r^5$ \cite{QS3}. 
We wish to further mention that it was shown
in Ref. \cite{SS2} that 
$W_x (z)$ deviates from $v_x (z)$ also in the metal bulk 
due to the contribution from $W_{t_C}^{(1)} (z)$. Only in the LDA,
$W_x (z)$ has the same bulk value of $-k_F/\pi$ as $v_x (z)$, as shown
in Ref. \cite{wang}. 
In this connection we list the long-known fact for $V_x^S$ $(z)$:
\begin{eqnarray}
V_x^S (z) = - \frac{3k_F}{ 2\pi} ~~~~~~~~~ for ~~ z = -\infty .
\end{eqnarray}
Therefore, strictly speaking, $V_x^{KLI} (z)$ only, among
all the approximate exchange potentials considered here, has the 
remarkable property with the same bulk value as that of $v_x (z)$.

Finally we would make several comments on the results reported
in Ref. \cite{horowitz2}. The large-distance structure of $v_x(z)$
was attributed solely to $V_x^\bigtriangleup (z)$ in Ref. \cite{horowitz2},
and consequently the conclusion $v_x (z) \sim \ln z/z$ was reached.
On the other side, $V_x^{shift} (z)$ was claimed to
decay as $\sim \ln z /z^2$, and hence make no
leading order contribution to $v_x (z)$ \cite{horowitz2}.
On the contrary, we find that in fact
\begin{eqnarray} \label{vsh1}
V_x^{shift} (z \to \infty) 
= && -\frac{\sqrt{\beta^2 -1}}{2 \pi}  \frac{1}{z}
\biggl [ \ln \biggl ( \frac{k_Fz}{\sqrt{\beta^2 -1}} 
\biggr )  \nonumber \\
&& + C + 2 \ln 2 -1 \biggr ]  
+ \alpha_x \frac{1}{z} .
\end{eqnarray}
The leading order of $O(\ln z/z)$ therefore exactly cancels out
in $v_x (z)$, resulting in a
faster decay of $O(1/z)$.
The discrepancy of $V_x^{KLI} (z)$ from $v_x (z)$ in the classically
forbidden region is also
due to the ignored contribution [cf. Eq. (\ref{vx-vkli})]
from $V_x^{shift} (z)$. 

\section{Asymptotic structures of $v_x (z)$, 
$V_x^S (z)$, $V_x^{KLI} (z)$, and $W_x (z)$
at the slab surface}

So far we have considered only the metal surface with the
semi-infinite geometry. 
A great deal of work of the electronic structure at the metal 
surface has been carried out on a jellium slab instead.
We shall consider the slab case in this section. To this end, 
we first mention two studies in which the asymptote 
of the xc potential in the classically forbidden region was
addressed as one of the key issues. One was 
reported in Ref. \cite{horowitz1}, mentioned previously in the
Introduction, in which it was claimed that $v_x (z \to \infty) 
= - 1 /z$ asymptotically. 
Numerical calculation in Ref. \cite{eguiluz} 
based on the GW approximation to the
electron exchange-correlation 
self-energy $\Sigma_{xc}$, however, yielded the different
result that $v_x (z) \sim
-1/z^2$ and $v_c (z)
\sim -1/(4z)$.  Both studies were performed on the 
slab surface.
Part of our effort in this section will be devoted to 
sheding some light on these results.
In fact it is found that 
the asymptotic behavior of $v_x (z)$, $V_x^S (z)$,
$V_x^{KLI} (z)$, and $W_x (z)$
all depends critically on the width 
of the slab.

We consider a metal slab with a typical width not 
exceedingly larger than $\lambda_F$ or, in other words, comparable 
to or smaller than $\lambda_F$ where $\lambda_F$ is the bulk Fermi
wavelength. 
In this case the discreteness of the eigenenergies of
the electron in the slab must be taken into account. The system can be
regarded virtually as a finite one, and the well-known conclusion for
the finite system that $v_x ({ \bf r}) \sim -1/r$ mentioned in Sec. III 
therefore holds. As a matter of fact, 
this already explains the result in Eq. (\ref{vxslab0}).
More explicitly, the Dirac density matrix has the following
well-known asymptotic form:
\begin{eqnarray}  \label{gammasm}
\gamma_s ({\bf r}, {\bf r}') = 2 \phi_m ({\bf r}) 
\phi_m^* ({\bf r}') ~~~~~ for ~~~z \to \infty.
\end{eqnarray} 
From Eq. (\ref{HF-pot}), by the use of 
multipole expansion argument for the Coulomb interaction 
which evidently is applicable here, one immediately has
\begin{eqnarray}  \label{uxm}
u_{xm} (z \to \infty) = -\frac{1}{z} .
\end{eqnarray}
The notation $m$ clearly has the same meaning as $k_F$ but
is used instead to emphasize
the discreteness of the eigenenergies.
Equation (\ref{vxslab0}) then follows from Eq. (\ref{ident}).

Additional care is needed in the above discussion for $v_x (z)$ 
in that it is not fully rigorous since actually $D_{k_F} \neq 0$
in the case of the slab metal surface. Equation (\ref{ident})
therefore is not valid and we in effect 
must resort to Eq. (\ref{ident0}) instead.
This point will get further refined near the end of this section 
[cf. Eq. (\ref{vxslab2}) below].

In any way, 
the fact that the planar freedom remains extending to infinity 
is essentially irrelevant in the present discussion. 
The key point is the dominant behavior of the highest occupied 
orbital and localization of the exchange hole near the finite system. 
The slab clearly fails to
catch the continuity feature of the electron eigenenergies 
which is essential to the semi-infinite
metal surface. Equation (\ref{vxslab0}) thus has limitation in that
it is valid only for the slab surface and can not be
naively extrapolated for the semi-infinite surface.
Nevertheless, a numerical demonstration of Eq. (\ref{vxslab0})
for the slab surface like that in Ref. \cite{horowitz1} 
is still quite valuable.

The effects of the delocalization of the exchange hole are 
negligible in the case of the slab surface.
$V_x^S (z)$ consequently has the same asymptotic form as $v_x (z)$:
\begin{eqnarray} \label{vxs55}
V_x^S ( z \to \infty)  = - \frac{1}{ z}
\end{eqnarray} 
at limiting large $z$. Correspondingly
$\epsilon_x$ $(z \to \infty)$ $= -\rho (z)/2z $. 
On the other hand $\rho (z)$ has the following asymptotic form:
\begin{eqnarray} 
\rho ( z \to \infty)  = 2 |\phi_{k_F} (z)|^2 .
\end{eqnarray}
The reader is referred to Eq. (\ref{density99}) for comparison.
Equation (\ref{vxs55}) can be 
readily obtained from Eq. (\ref{slater-def}) via keeping the 
leading-order term in the
multipole expansion for the Coulomb interaction. It is basically just
another illustration of the well-known result 
$V_x^S ( {\bf r} \to \infty) =-1/r$ for a finite system.
Indeed, it follows from Eqs. (\ref{exchangehole}) and (\ref{gammasm}) 
that, for $z \to \infty$,
\begin{eqnarray} \label{rhoxas}
\rho_x ({\bf r}, {\bf r}') = |\phi_m ({\bf r}')|^2 .
\end{eqnarray}
Equation (\ref{vxs55}) immediately follows from Eqs. (\ref{slater-def})
and (\ref{rhoxas}). Similarly, Eq. (\ref{ex-curl}) together with
Eq. (\ref{rhoxas}) yields
\begin{eqnarray}
{\bf \cal E}_x ({\bf r}) = - \frac{1}{z^2} {\bf e}_z ,
\end{eqnarray}
and it then follows from Eq. (\ref{wx0}) that
\begin{eqnarray} 
W_x ( z)  = - \frac{1}{ z} .
\end{eqnarray}
$V_x^{KLI} (z)$ possesses 
the same asymptotic form [but see also Eq. (\ref{KLIslab})].
Only in this case $V_x^{shift} (z)$,
which now decays exponentially at large $z$, 
is much smaller than $v_x^{\bigtriangleup}
(z)$ which in contrast approaches a nonzero constant:
\begin{eqnarray}  \label{Dm5}
v_x^{\bigtriangleup} (z) = D_{m}   ~~~~~~~~~for  ~~~~ z \to \infty .
\end{eqnarray}
We note once again that $k_F$ has the same meaning as $m$. The fact 
that $D_{\bf k}|_{|{\bf k}|=k_F}$ $=D_{k_F}$  guarantees no ambiguity 
in the meaning of $D_m$.

In summary Eq. (\ref{uxm}) [or Eq. (\ref{vxs55})] can be
obtained for the slab case via the argument 
for the multipole expansion of the Coulomb interaction. The
metal-surface feature of the slab plays no crucial role at this point.
Therefore it is not necessary to 
resort to detailed derivations based
on Eq. (\ref{A30}) [or Eq. (\ref{vslat0})]. [Equations (\ref{A30}) and
(\ref{vslat0}) hold for both the cases of the semi-infinite
and the slab metal surfaces.] Nevertheless 
such derivations turn out to be amazingly simple (due of course 
also to the feature of the finiteness of the system). In the
meanwhile they could
be fairly illuminating. We therefore include one
in the following mainly for the purpose of 
illustration. To this end, we first 
copy Eq. (\ref{A30}) below with appropriate
modifications in the form for the slab case:
\begin{eqnarray} \label{uxk44}
u_{x k_F} (z)= &&-\frac{2}{L \phi_{k_F} (z)}
\sum_k^{occ.} \lambda \phi_k (z) \nonumber \\
&&\int_{-L}^0 d z' \phi_k^* (z')
\phi_{k_F} (z') \nonumber \\
&&\int_0^\infty dr_{\parallel}' \frac{1}
{\sqrt{(z -z')^2 +r_{\parallel}'^2}} J_1 (\lambda r_{\parallel}') .
\end{eqnarray}
Only the main domain ($-L \le z' \le 0$) has been taken into 
account for the integral over $z'$, which is clearly justified. 
Since $z >> z'$, the denominator 
$\sqrt{(z- z')^2 + r_{\parallel}'^2}$ can be readily replaced
by $\sqrt{z^2 + r_{\parallel}'^2}$. The dominant
contribution to the integration over $k$ arises from the
region of $k_F -k  \sim 1/z$, and accordingly $\lambda >> k_F - k$.
On the other hand, The dominant contribution to the integration over
$r_{\parallel}'$ arises from the region of
$ 0 \le $ $ r_{\parallel}'$ $\alt O(\lambda^{-1})$
since $ J_1 (\infty) =0$. Correspondingly $r_{\parallel}' << z$
and hence $\sqrt{z^2 + r_{\parallel}'^2}$ can be further replaced
by $z$. The integration over $r_{\parallel}'$ then 
turns out simply to be
\begin{eqnarray}
\int_0^\infty dr_{\parallel}' J_1 (\lambda r_{\parallel}')
=\frac{1}{ \lambda} .
\end{eqnarray}
Equation (\ref{uxk44}) as a consequence becomes
\begin{eqnarray} 
u_{x k_F} (z \to \infty)= &&-\frac{2}{z L \phi_{k_F} (z)}
\sum_k^{occ.} \phi_k (z) \nonumber \\
&&\int_{-L}^0 d z' \phi_k^* (z')
\phi_{k_F} (z') .
\end{eqnarray}
One then employs the following equation:
\begin{eqnarray}
\int_{-L}^0 d z' \phi_k^* (z')
\phi_{k_F} (z') = \frac{L}{2} \delta_{k, k_F} .
\end{eqnarray}
Equation (\ref{uxm}) follows (with $m$ equivalent to $k_F$). 
Equation (\ref{vxs55}) can be obtained
from Eq. (\ref{vslat0}) in a similar manner.

Notice that, strictly speaking, 
\begin{eqnarray} \label{KLIslab}
V_x^{KLI} (z \to \infty)= D_m -\frac{1}{z} ,
\end{eqnarray}
where the term of the constant $D_m$ arises from the 
$V^\bigtriangleup _x (z)$ component on the rhs of Eq. (\ref{kli}), as 
shown in Eq. (\ref{Dm5}). Furthermore, in accordance to the claims made 
for finite systems in Ref. \cite{gorling},
such type of constant might also possibly occur in $v_x (z)$, since as
mentioned above the slab can be regarded essentially as a finite system.
Indeed, since $D_{k_F} \neq 0$ for the slab surface, one should 
resort to Eq. (\ref{ident0}) instead of Eq. (\ref{ident}). 
Equation (\ref{ident0}) in fact can be understood as one of 
the special cases of the following general result:
\begin{eqnarray} \label{uxm+Dm} 
v_x ({\bf r}) = u_{xm} ({\bf r}) + D_m ,
\end{eqnarray}
proposed for the finite system in Ref. \cite{gorling}.
In the light of Eq. (\ref{uxm}) we consequently have 
\begin{eqnarray} \label{vxslab2}
v_x (z \to \infty) =  D_{m} - 1/z,
\end{eqnarray}
instead of Eq. (\ref{vxslab0}). Therefore, strictly speaking,
$v_x(\infty)$ $\neq 0$ now, [see also the discussions in 
the next paragraph.] Equation (\ref{vxslab2}) can be
alternatively obtained from Eq. (\ref{vx-vkli}) and Eq. (\ref{KLIslab}),
together with the fact that $V_x^{shift} (z \to \infty)$ vanishes 
exponentially. 

Finally we remark that $V_x^S (\infty)$ $=0$ in any case. On the other
hand, Eq. (\ref{wx0}) automatically gurantees $W_x (\infty ) =0$. 
Furthermore, $D_m$ can be made to vanish so that $v_x (\infty)=0$ 
by shifting $v_x ({\bf r})$ in the finite system \cite{gorling} 
or the slab of the present case. It is exactly in this sense 
that we justify
the result in Eq. (\ref{vxslab0}). However we wish also to mention that 
it remains unclear whether $v_x (\infty)$
can always be simultaneously shifted to be zero in the 
exact (full) KS scheme in which one has already adopted $v_{xc}(\infty)=0$.
The remarkable fact is that $D_m$ in Eq. (\ref{vxslab2})
evolves into $D_{\bf k}$ at $|{\bf k}|=k_F$
of Eq. (\ref{Dk-vect}) and vanishes 
for the semi-infinite metal surface. Thus for 
the semi-infinite metal surface one has definitely $v_{xc} (\infty) =0$
and $v_x (\infty) =0$.

\section{Conclusions}

By the use of the OEP method, we have established an identity 
between the planar-momentum averaged orbital-dependent 
exchange potential $u_{xk_F} (z)$ and
the KS exchange potential  $v_x (z)$ in the classically forbidden region 
of the metal surface. Based on it, the asymptotic form of $v_x (z)$ 
at large distance from the metal surface has been investigated. 
The result is $v_x (z \to \infty) 
= - \alpha_x/z$, which 
%confirms the same result obtained before, and hence also 
resolves the controversy raised recently in the literature.
The point that $v_{xc} (\infty)=0$ and $v_x (\infty) =0$ hold simultaneously 
gets emphasized and carefully elucidated.
The asymptotic form of the Slater exchange potential $V_x^S (z \to \infty)
=-2 \alpha_x/z$ is also verified.
The result for $v_x (z)$ and that for $V_x^S (z)$ were initially proposed
in Ref. \cite{SS1}, and the former was verified
subsequently in Ref. \cite{QS1}.
The further confirmation in the present work indicates beyond doubt
that they are correct and the issue is finally settled.
Furthermore, the structure of the exchange hole
has been examined, and especially the delocalization 
nature of it 
for an electron far outside the metal surface 
has been demonstrated.
It is exactly such delocalization that gives rise to the quite nontrivial
asymptotic behavior of $v_x (z)$. In addition, the asymptotic structure of 
the approximate KLI exchange potential $V_x^{KLI} (z)$ 
and HS exchange potential $W_x (z)$ at large $z$ has also been 
surveyed, which are of the
forms $V_x^{KLI} (z \to \infty) \sim \ln z/z$ 
and $W_x (z \to \infty) \sim -\alpha_W /z$, respectively. 

As mentioned in the Introduction, common wisdom favors the belief 
that the full Kohn-Sham
exchange-correlation potential $v_{xc} (z)$ decays like the classical
image potential at large distance from the 
metal surface. Doubt however has been casted on it 
in Ref. \cite{QS1} in which the asymptote of $v_c (z)$ was also 
studied. It was shown in Ref. \cite{QS1} that the asymptotic form
of $v_{xc} (z)$ is not the same as 
the classical image potential.
Recent progress on the asymptote of $v_{c} (z)$ has 
also been made in Ref. \cite{constantin}. Unfortunately approximations
have been employed in the calculations for $v_{c} (z)$ in both of
Refs. \cite{QS1} and \cite{constantin} and it is not clear whether
they are fully justified. 
The subject of the asymptote of $v_{xc} (z)$ thus 
remains not fully settled.

The asymptotic behavior of
the exchange potential at the metal slab surface has also been
investigated. It is shown that
asymptotically $v_x (z)$, as well as $V_x^S (z)$, 
$V_x^{KLI} (z)$, and $W_x (z)$,
depends critically on the width of the slab.
In particular, if the width 
is comparable to or smaller than $\lambda_F$, the slab can be essentially
regarded as a finite system and asymptotically $v_x (z) = -1/z$.
$V_x^S (z)$, 
$V_x^{KLI} (z)$, and $W_x (z)$ all have this same form.
The exchange energy density $\epsilon_x (z)$ correspondingly approaches 
asymptotically $-\rho (z)/2z$. 
All these facts are inherently due to the localization of the exchange hole
in the finite system. While by definition $V_x^S (z)$ and $W_x (z)$
vanish in the classically forbidden region of both the semi-infinite 
and the slab metal surfaces, a careful analysis reveals 
that the $-1/z$ term commences only to the second order contribution 
to $v_x (z)$ and $V_x^{KLI} (z)$
and actually both $v_x (z \to \infty)$ and $V_x^{KLI} (z \to \infty)$
approach a nonzero constant, i.e., $v_x (\infty)$ $=V_x^{KLI} (\infty)$
$=D_m$ $\neq 0$. The constant $D_m$ however can always be made to vanish in th
exchange-only self-consistent calculations by shifting $v_x (z)$. It
is precisely in this sense that
one has the result of Eq. (\ref{vxslab0}).
% $v_x (z \to \infty) = -1/z$ and
%$V_x^{KLI} (z \to \infty) = -1/z$, the same form as
%$V_x^S (z)$ and $W_x (z)$. 

This work was supported by the National Science Foundation of China.

\appendix

\section{Proof for Eq. (\ref{ur=uz})}

We substitute Eq. (\ref{A-gamma}), together with Eq. (\ref{orbital}),
into Eq. (\ref{HF-pot}). This results in
\begin{eqnarray}
u_{x {\bf k}} ({\bf r}) = && - \frac{1}{ \pi^2 \phi_{\bf k} ({\bf r})}
\sqrt{\frac{2}{AL}} \int_0^{k_F} dk' \lambda ' \phi_{k'} (z) \nonumber \\
&& \int_{-\infty}^\infty
dz' \phi_{k'}^* (z') \phi_k (z')  \nonumber \\
&& \int d {\bf x}'_\parallel e^{i{\bf k}_\parallel 
\cdot {\bf x}'_\parallel}
\frac{J_1 (\lambda ' |{\bf x}_\parallel - {\bf x}'_\parallel|)}
{|{\bf r} - {\bf r}'| |{\bf x}_\parallel - {\bf x}'_\parallel|} .
\end{eqnarray}
%where $\lambda ' = \sqrt{k_F^2 -k'^2}$.
Making the transform ${\bf r}'_\parallel=
{\bf x}_\parallel - {\bf x}'_\parallel$ in the above equation  
and making the use 
of Eq. (\ref{orbital}) once again, we establish Eq. (\ref{ur=uz}) 
explicitly with the following expression:
\begin{eqnarray} 
u_{x {\bf k}} (z) = && - \frac{1}{\pi^2 \phi_k (z)}
\int_0^{k_F} dk' \lambda' \phi_{k'} (z) \nonumber \\
&& \int_{-\infty}^\infty
dz' \phi_{k'}^* (z') \phi_k (z')    \nonumber \\
&& \int d {\bf r}'_\parallel e^{i{\bf k}_\parallel 
\cdot {\bf r}'_\parallel}
\frac{J_1 (\lambda ' r'_\parallel)}
{r'_\parallel \sqrt{(z-{z'})^2 + {r'}_\parallel^2} } .
\end{eqnarray}
Further algebra then yields
\begin{eqnarray} \label{ux5}
u_{x {\bf k}} (z) = && - \frac{2}{\pi \phi_k (z)}
\int_0^{k_F} dk' \lambda' \phi_{k'} (z) \nonumber \\
&& \int_{-\infty}^\infty
dz' \phi_{k'}^* (z') \phi_k (z')    \nonumber \\
&& \int_0^\infty d r'_\parallel 
\frac{J_0 (k_\parallel r'_\parallel) J_1 (\lambda ' r'_\parallel)}
{\sqrt{(z-{z'})^2 + {r'}_\parallel^2} } . 
\end{eqnarray}

\section{Derivation for Eq. (\ref{uxkf})}

It is not difficult to see that, for limiting large $z$, 
the leading contribution to the integral over $k$ on 
the rhs of Eq. (\ref{A30}) arises from the 
region $k_F - k \sim 1/z$, and that to 
the integral over $z'$, on the other side,
from the region of the metal bulk. 
Accordingly, $\lambda >> (k_F -k)$ 
and hence $|z -z'| >> 1/\lambda$. Since $J_1 (\infty) = 0$ and 
accordingly the dominant contribution to the integral 
over $r_{\parallel}'$ arises from the region of $0 \le $ $r_{\parallel}'$
$\alt O(\lambda^{-1})$, one has $|z - z'| >> r_{\parallel}'$ in the
integral over $r_{\parallel}'$ in Eq. (\ref{A30}).
Thus it follows that
\begin{eqnarray}
\int_0^\infty dr_{\parallel}' &&\frac{1}
{\sqrt{(z -z')^2 +r_{\parallel}'^2}} J_1 
(\lambda r_{\parallel}') \nonumber \\
&& = \frac{1}{|z - z'|} \int_0^\infty
dr_{\parallel}' 
J_1 (\lambda r_{\parallel}') \nonumber \\
&& = \frac{1}{\lambda |z - z'|}. 
\end{eqnarray}
Correspondingly,   
\begin{eqnarray}  \label{ukF}
u_{x k_F} (z \to \infty) &&= -\frac{2}{\pi \phi_{k_F} (z)}
\int_0^{k_F} dk  \phi_k (z)   \nonumber \\
&& \int_{-\infty}^{\infty} d z' \frac{1}{|z - z'|}  \phi_k^* (z')
\phi_{k_F} (z') . 
\end{eqnarray}
We next define
\begin{eqnarray} \label{theta}
\Theta_{k, k'} (z) = 2 
\int_{-\infty}^{\infty} d z' \frac{1}{|z - z'|}  \phi_{k'}^* (z')
\phi_{k} (z') .
\end{eqnarray}
Since, as just mentioned, the integral over $z'$ arises mainly from 
the bulk region, one can make the use of Eq. (\ref{sine}) 
for the orbitals in the above expression and it follows that 
\begin{eqnarray} \label{te5}
\Theta_{k, k'} (z) =  
\int_{-\infty}^{-d} \frac{d z'}{ z-z'} &&
[\cos ( k_{-}z' +  \delta_{-}) \nonumber \\ 
&&-\cos ( k_{+} z' +  \delta_{+}) ] ,
\end{eqnarray}
where $ k_{\pm} =k \pm k'$, $ \delta_{\pm} =\delta (k) \pm \delta (k')$,
and $-d$ stands for a negative position near the surface whose exact
value is irrelevant for $z >> d$. Indeed, 
\begin{eqnarray} 
\Theta_{k, k'} (z) =
\int_{-\infty}^{z+d} \frac{d z'}{ z'} &&
[\cos \{ k_{-} (z-z') +  \delta_{-}) \} \nonumber \\
- && \cos \{ ( k_{+} (z-z') +  \delta_{+}) \}] ,
\end{eqnarray}
and the upper limit of the integral: $z+d$, can be readily replaced by $z$.
Thus one has
\begin{eqnarray} \label{te6}
\Theta_{k, k'} (z) &&=  \cos ( k_{-}z
+ \delta_{-}) \int_z^\infty \frac{d z'}{z'} 
\cos  (k_{-}z')  \nonumber \\
&& + \sin ( k_{-}z
+  \delta_{-}) \int_z^\infty \frac{d z'}{z'} 
\sin  (k_{-}z')  \nonumber \\
&& - \cos ( k_{+}z
+  \delta_{+}) \int_z^\infty \frac{d z'}{z'}
\cos  (k_{+}z')  \nonumber \\
&& - \sin ( k_{+}z
+  \delta_{+}) \int_z^\infty \frac{d z'}{z'}
\sin  (k_{+}z') , 
\end{eqnarray}
or
\begin{eqnarray} \label{te55}
\Theta_{k, k'} (z)  =&&    -\cos ( k_{-} z
+  \delta_{-}) ci ( k_{-}z)  \nonumber \\
 && -  \sin ( k_{-}z
+  \delta_{-})  si ( k_{-}z) \nonumber \\
&&    + \cos ( k_{+} z
+  \delta_{+}) ci ( k_{+}z)  \nonumber \\
 && +  \sin ( k_{+}z
+  \delta_{+})  si ( k_{+}z),
\end{eqnarray}
where $si $ and $ ci $ are the sine integral and cosine
integral, respectively \cite{gradshteyn}:
\begin{eqnarray}
si (x) = - \int_x^\infty \frac{ dt}{t} \sin t ,
\end{eqnarray}
\begin{eqnarray}
ci (x) = - \int_x^\infty \frac{ dt}{t} \cos t .
\end{eqnarray}
For $z \to \infty$ the third and fourth terms on the rhs of
Eq. (\ref{te55}) will be demonstrated at the end of this appendix to 
make only higher-order contribution to $u_{xk_F} (z)$. Consequently
these terms can be ignored. Hence one comes to
\begin{eqnarray} \label{theta2}
\Theta_{k, k'} (z \to \infty)  =&&    -\cos ( k_{-} z
+  \delta_{-}) ci ( k_{-}z)  \nonumber \\
 && -  \sin ( k_{-}z
+  \delta_{-})  si ( k_{-}z) ,  
\end{eqnarray}
We now substitute Eqs. (\ref{theta}) and (\ref{theta2}) 
(with $k$ replaced by $k_F$, and $k'$ by $k$) 
into Eq. (\ref{ukF}), and obtain
\begin{eqnarray}
u_{x k_F} (z \to \infty) &&= \frac{1}{\pi \phi_{k_F} (z)}
  \int_0^{k_F} dk \phi_k (z)   \nonumber \\
&& [\cos (a +  {\tilde \delta}_{-}) ci (a) 
 + \sin (a + {\tilde \delta}_{-}) si (a)],  \nonumber \\
\end{eqnarray}
where 
${\tilde \delta}_{-} = \delta(k_F) - \delta(k)$, 
and $a = (k_F - k)z$. Again, as mentioned above,
for large $z$ the dominant contribution to the integral in
the above equation arises from the region $k_F -k \sim 1/z$.
Correspondinly ${\tilde \delta}_{-} \to 0$ and  $\phi_k (z)$ 
$=\phi_{k_F} (z)$ $e^{-ca}$ where $c= 1/\sqrt{\beta^2 -1}$. 
Therefore we finally obtain Eq. (\ref{uxkf}), with
\begin{eqnarray} \label{A-alphax}
\alpha_x = - \frac{1}{\pi} \int_0^\infty
da e^{-ca} [  ci (a) \cos a + si (a) \sin a ] .
\end{eqnarray}
The value of $\alpha_x$ given in Eq. (\ref{alphax}) may
be obtained via carrying out the integral 
in Eq. (\ref{A-alphax}) \cite{gradshteyn}:
\begin{eqnarray}
\int_0^\infty da e^{-ca} [  ci (a) \cos a + si (a) \sin a ]
=-\frac{\pi + 2 c  \ln c}{2(1 +c ^2)} .   \nonumber \\
\end{eqnarray}
Solely to draw a connection with the calculations performed in an
alternate approach in Ref. \cite{QS1}, we mention the following
identity \cite{gradshteyn}:
\begin{eqnarray} 
\int_0^\infty du \frac{u}{u^2 + a^2} e^{-u}  
=  - ci (a) \cos a - si (a) \sin a .
\end{eqnarray}  

Were the third and fourth terms in Eq. (\ref{te55}) taken into account,
they would have made the following contribution to $u_{x k_F} (z)$:
\begin{eqnarray}
-\frac{1}{ c \pi z}  [ && \cos \{2k_F z + \delta(k_F) \}  
ci (2k_F z) \nonumber \\
&& + \sin \{2k_F z + \delta(k_F) \}  si (2k_F z) ] .
\end{eqnarray} 
Since $ci (2k_F z)$ $\to 0$ and $si (2k_F z)$ $\to 0$ as $z \to \infty$, 
this contribution is therefore of the order $o(1/z)$ and the neglect 
of these terms in Eq. (\ref{te55}) is thus justified.

\section{Verification of Eq. (\ref{Vslater}) }

The $\phi_k (z)$ is in fact real, and hence the integrals
over $k$ and $k'$ in Eq. (\ref{vslat0}) are clearly symmetric. 
Accordingly we change the domain
of the $k'$ integration in Eq. (\ref{vslat0}) 
to $\int_0^k d k'$ combined with a corresponding double.
Similar arguments to those at the beginning of Appendix B then lead,
for $z \to \infty$, to
\begin{eqnarray} \label{te7}
\int_0^\infty && dr_{\parallel}' \frac{1}
{r_{\parallel}'
\sqrt{(z -z')^2 +r_{\parallel}'^2}} J_1 (\lambda r_{\parallel}') 
J_1 (\lambda' r_{\parallel}')  \nonumber \\
&&= \frac{1}{ |z -z'|} \int_0^\infty  d t \frac{1}
{t} J_1 (t)
J_1 (\lambda' t/\lambda)  \nonumber \\
&& = \frac{\lambda }{2 \lambda' |z -z'|} .  
\end{eqnarray}
Thus it follows from Eq. (\ref{vslat0}) that, for $z \to \infty$,
\begin{eqnarray}    \label{Vslater5}
V_x^S (z) = - && \frac{4}{\pi^3 \rho (z)} 
\int_0^{k_F} dk \phi_k (z) \lambda^2 \int_0^{k} dk' 
\phi_{k'}^* (z) \nonumber \\
&& \int_{-\infty}^{\infty} d z' \frac{1}{|z - z'|}  
\phi_{k'}^* (z') \phi_{k} (z') .
\end{eqnarray}
Making the use of Eq. (\ref{theta}) in the 
preceding equation yields, for $ z \to \infty$, further
\begin{eqnarray}  \label{slater4}
V_x^S (z) = - \frac{2}{\pi^3 \rho (z)} && 
\int_0^{k_F} dk \phi_k (z)  \lambda^2 \nonumber \\
&& \int_0^{k} 
dk' \phi_{k'}^* (z) \Theta_{k', k} (z).
\end{eqnarray}
Clearly the third and fourth terms in Eq. (\ref{te55}) for
$\Theta_{k', k} (z)$ can be ignored once again.  Accordingly
Eq. (\ref{theta2}) for $\Theta_{k', k} (z)$ is substituted into 
Eq. (\ref{slater4}) instead. We make
further the transform of the integral variable $k'=k -b/z$. 
It follows then that, for $z \to \infty$,
\begin{eqnarray}  
V_x^S (z) &&=  \frac{2}{\pi^3 \rho (z)} 
\int_0^{k_F} dk |\phi_k (z)|^2 \lambda^2   \nonumber \\
&& \int_0^\infty 
db e^{-cb} [  ci (b) \cos b + si (b) \sin b ] .
\end{eqnarray} 
We next compare the above equation with Eq. (\ref{A-alphax}) to 
obtain Eq. (\ref{Vslater}). In doing so the expression of (\ref{den5}) 
for $\rho (z)$ has also been employed.

\section{Derivations for Eq. (\ref{rhox2}), and 
Eqs. (\ref{bx1}) and (\ref{bx2})}

For limiting large $z$, the leading-order contribution
to the integral over $k'$ on the rhs of Eq. (\ref{twoterm}) 
arises from the region of
$k' \to k$, and hence we have
\begin{eqnarray}
\rho_x (z, z')= -\frac{4}{\pi^3 z\rho(z)}
\int_0^{k_F} dk && \lambda^2 \frac{\kappa}{k}
\phi_k^* (z) \phi_k (z')   \nonumber \\
&& \phi_{k} (z) \phi_{k}^* (z') .
\end{eqnarray}
%Analogously,
Similarly the contribution to the integral over $k$ arises from
the region of $k \to k_F$. Therefore
\begin{eqnarray}
\rho_x (z, z')=  -\frac{4}{\pi^3 z\rho(z)}\frac{1}{c}
 |\phi_{k_F} (z')|^2  
\int_0^{k_F} dk \lambda^2 |\phi_k (z)|^2 .   \nonumber \\
\end{eqnarray}
We then make the use of the expression (\ref{den5}) for $\rho (z)$. 
Equation (\ref{rhox2}) follows.

We next present the derivation leading to Eqs. (\ref{bx1}) 
and (\ref{bx2}). The explicit expression for $F(q)$ has been reported
in Refs. \cite{SS1,QS1} as
\begin{eqnarray}  \label{F(q)}
F(q) && = \pi \lambda_<^2 \theta(\lambda_> - \lambda_< - q)  \nonumber \\
&& + \theta(\lambda_> + \lambda_<
-q) \theta(q-\lambda_> + \lambda_<)  \nonumber \\
&& \times [\pi \lambda_<^2 \
\theta \{ (\lambda_>^2 - \lambda_<^2 )^{1/2} -q \}
+ S_{\lambda_>} (q) + S_{\lambda_<} (q) ] ,  \nonumber \\
\end{eqnarray}
where $\lambda_>$ ($\lambda_<$) is the larger (smaller) one
of $\lambda$ and $\lambda '$, respectively.
In Eq. (\ref{F(q)}),
\begin{eqnarray}  \label{S(q)}
S_{\lambda} (q) = \lambda^2 tan^{-1} \frac{(\lambda^2 -
X_{\lambda}^2)^{1/2} }{X_{\lambda}}
- X_{\lambda} ( \lambda^2 - X_{\lambda}^2 )^{1/2} ,    \nonumber \\
\end{eqnarray}
\begin{eqnarray}
S_{\lambda'} (q) = S_{\lambda} (q) |_{\lambda \rightarrow \lambda'} ,
\end{eqnarray}
and
\begin{eqnarray}
X_{\lambda} (q) =\frac{1}{2} (q +
\frac{\lambda^2 - \lambda'^2}{q^2} ) ;
\end{eqnarray}
\begin{eqnarray}
X_{\lambda'} (q) =\frac{1}{2} (q -
\frac{\lambda^2 - \lambda'^2}{q^2} ) .
\end{eqnarray}
The fact that the 
integral $\int_0^\infty dq F(q)$ can be carried out analytically is the
key point in obtaining the neat final expressions of (\ref{bx1}) 
and (\ref{bx2}) for $b_x (z, z')$.
To this end, we write it as
\begin{eqnarray} 
\int_0^\infty dq F(q) && =  \int_0^{\sqrt{\lambda_>^2 - \lambda_<^2}}
dq \pi \lambda_<^2  \nonumber \\
 + && \int_{\lambda_> - \lambda_<}^{\lambda_> + \lambda_<}
dq [ S_{\lambda_>} (q) + S_{\lambda_<} (q)] .
\end{eqnarray}
The first integral on the rhs of the above equation 
is trivial. After performing a partial integration over $q$
in the second one, we come to
\begin{eqnarray} \label{integ55} 
\int_0^\infty dq F(q)= -
\int_{\lambda_> - \lambda_<}^{\lambda_> + \lambda_<}
dq \frac{\partial}{\partial q} 
[S_{\lambda_>} (q) + S_{\lambda_<} (q)] .   \nonumber \\
\end{eqnarray}
We caution the reader that
the function $S_{\lambda_>} (q) + S_{\lambda_<} (q)$ has a
discontinuity at $q= \sqrt{\lambda_>^2
-\lambda_<^2}$:
\begin{eqnarray}
S_{\lambda_>} (\sqrt{\lambda_>^2
-\lambda_<^2}_{-}) + S_{\lambda_<} (\sqrt{\lambda_>^2
-\lambda_<^2}_{-})= -\frac{\pi}
{2} \lambda_<^2, \nonumber \\
\end{eqnarray}
\begin{eqnarray}
S_{\lambda_>} (\sqrt{\lambda_>^2
-\lambda_<^2}_{+}) + S_{\lambda_<} (\sqrt{\lambda_>^2
-\lambda_<^2}_{+})= \frac{\pi}
{2} \lambda_<^2 .  \nonumber \\
\end{eqnarray}
Care therefore is needed in obtaining Eq. (\ref{integ55}).
We next list the following properties: 
\begin{eqnarray} 
\lambda_>^2 - X_{\lambda_>}^2 = \lambda_<^2 - X_{\lambda_<}^2 ,
\end{eqnarray}
and
\begin{eqnarray} 
\frac{\partial}{\partial q} S_{\lambda_>} (q) = 
\frac{\partial}{\partial q}  S_{\lambda_<} (q)
=- \sqrt{\lambda_<^2 - X_{\lambda_<}^2} . 
\end{eqnarray}
They are found to be useful in our further simplifying Eq. (\ref{integ55}) to
\begin{eqnarray} \label{integ0}
\int_0^\infty dq F(q)= 2
\int_{\lambda_> - \lambda_<}^{\lambda_> + \lambda_<}
dq q ( \lambda_<^2 - X_{\lambda_<}^2 )^{1/2} .
\end{eqnarray}
Equation (\ref{integ0}) is rewritten trivially as
\begin{eqnarray} \label{integ5}
\int_0^\infty dq F(q)  && = 2
\int_{\lambda_> - \lambda_<}^{\sqrt{\lambda_>^2 - \lambda_<^2}}
dq q ( \lambda_<^2 - X_{\lambda_<}^2 )^{1/2}  \nonumber \\
 + && 2 \int_{\sqrt{\lambda_>^2 - \lambda_<^2}}^{\lambda_> + \lambda_<}
dq q ( \lambda_<^2 - X_{\lambda_<}^2 )^{1/2} .
\end{eqnarray}
The rhs of Eq. (\ref{integ5}) can be further simplified via the following 
transform in the integrals:
\begin{eqnarray}
x=\frac{1}{\lambda_<} ( \lambda_<^2 - X_{\lambda_<}^2 )^{1/2}.
\end{eqnarray}
Under this transform, 
\begin{eqnarray}
&&X_{\lambda_<} = - \lambda_< \sqrt{1 -x^2}, \nonumber \\
&&q = \sqrt{\lambda_>^2 -\lambda_<^2 x^2} -  
\lambda_< \sqrt{1 -x^2}  
\end{eqnarray}
in the first integral on the rhs of Eq. (\ref{integ5}); but 
\begin{eqnarray}
&&X_{\lambda_<} =  \lambda_< \sqrt{1 -x^2},  \nonumber \\
&&q = \sqrt{\lambda_>^2 -\lambda_<^2 x^2} + 
\lambda_< \sqrt{1 -x^2}   
\end{eqnarray}
in the second one. The resultant expression is
\begin{eqnarray} \label{integ}
\int_0^\infty dq F(q) = 4 \lambda_<^2 \int_0^1 
dx x^2 \frac{\lambda_>^2 + \lambda_<^2 (1 - 2 x^2)}{\sqrt{(\lambda_>^2
-\lambda_<^2 x^2)(1 - x^2)}}.  \nonumber \\
\end{eqnarray}
The preceding integral could be efficiently evaluated numerically. 
We here prefer to express it 
in terms of complete elliptic integrals. To this end,
we cite the following relations ($t \ge 1$) \cite{gradshteyn}:
\begin{eqnarray}
\int_0^1
dx x^2 && \frac{1}{\sqrt{(t^2 -x^2)(1 - x^2)}} \nonumber \\
&& = t \biggl [ F \biggl ( \frac{\pi}{2}, \frac{1}{t}
\biggr ) - E \biggl ( \frac{\pi}{2}, \frac{1}{t}
\biggr ) \biggr ]  \nonumber \\
&& = t \biggl [ {\bf K} \biggl ( \frac{1}{t}
\biggr ) -{\bf E} \biggl ( \frac{1}{t}
\biggr ) \biggr ] ,   \nonumber \\
\end{eqnarray}
and 
\begin{eqnarray}
\int_0^1 &&
dx  x^4 \frac{1}{\sqrt{(t^2 -x^2)
(1 - x^2)}}   \nonumber \\
&& =  \frac{1}{3} t \biggl [ (2t^2 +1)
F \biggl (\frac{\pi}{2},  \frac{1}{t}
\biggr ) - 2 (t^2 +1) E \biggl (\frac{\pi}{2}, \frac{1}{t}
\biggr ) \biggr ] \nonumber \\
&& =  \frac{1}{3} t \biggl [ (2t^2 +1) 
{\bf K} \biggl ( \frac{1}{t}
\biggr ) - 2 (t^2 +1) {\bf E} \biggl ( \frac{1}{t}
\biggr ) \biggr ] ,
\end{eqnarray}
where $F$ and $E$ are the elliptic integrals of the first 
and the second kinds, respectively.
Substituting these relations into Eq. (\ref{integ}), one has 
\begin{eqnarray} \label{b6}
\int_0^\infty dq F(q)=\frac{4 \lambda_>}{3} \biggl [
&& {\bf K} \biggl (\frac{\lambda_<}{\lambda_>} \biggr )
(\lambda_<^2 - \lambda_>^2)   \nonumber \\
+ && {\bf E} \biggl (\frac{\lambda_<}{\lambda_>} \biggr )
(\lambda_<^2 + \lambda_>^2) \biggr ].  
\end{eqnarray}
Further substitution of Eq. (\ref{b6}) into Eq. (\ref{b4}) then leads to
\begin{eqnarray}
b_x (z, z') = &&-\frac{4}{3 \pi^5 \rho (z) }
\int_0^{k_F} dk \int_k^{k_F} dk'   \nonumber \\
&& [\Phi_{k,k'} (z, z') + \Phi_{k',k} (z, z') ]  \nonumber \\
&&\lambda \biggl [
 {\bf K} \biggl (\frac{\lambda'}{\lambda} \biggr )
(\lambda'^2 - \lambda^2)
+{\bf E} \biggl (\frac{\lambda'}{\lambda} \biggr )
(\lambda^2 + \lambda'^2) \biggr ] .   \nonumber \\
\end{eqnarray}
Equation (\ref{bx1}) follows from the symmetry property 
$\Phi_{k,k'}$ $(z, z')$ $= \Phi_{k',k}$ $(z', z)$.

We next make the variable transform $y=\lambda'^2/\lambda^2$
in Eq. (\ref{G}) to
obtain, for $z \to \infty$,
\begin{eqnarray}
G(z, z') =
\phi_{k_F} (z) |\phi_{k_F} (z')|^2 
\int_0^{k_F} dk \lambda^5 \phi_k^*(z)  \nonumber \\
\int_0^1 \frac{dy}{2 \sqrt{k_F^2 - \lambda^2 y}}
e^{-c(k_F - \sqrt{k_F^2 - \lambda^2 y})z} \nonumber \\
\times [{\bf K}(\sqrt{y}) (y-1) + {\bf E} (\sqrt{y}) (y +1)].
\end{eqnarray}
Further transform $x=(k_F-k)z$ then yields
\begin{eqnarray} \label{G4}
G(z \to \infty, z') = && \frac{3 \pi^3}{16} 
k_F^{3/2} \gamma (cz)^{-7/2} \nonumber \\
&& |\phi_{k_F} (z)|^2 |\phi_{k_F} (z')|^2,
\end{eqnarray}
where
\begin{eqnarray} \label{gamma00}
\gamma = && \frac{32 \sqrt{2}}{3 \pi^3} c^{7/2}
\int_0^\infty dx \int_0^1 dy
e^{-cx(y+1)} x^{5/2}  \nonumber \\
&& \times [{\bf K}(\sqrt{y}) (y-1) + {\bf E} (\sqrt{y}) (y +1)].
\end{eqnarray}
Finally we substitute Eq. (\ref{G4}) into Eq. (\ref{bx1}) and 
apply the identity:
\begin{eqnarray}
\int_0^\infty dx x^{5/2} e^{-c(y+1)x} = \frac{15}{8} \sqrt{\pi}
[c(y+1)]^{-7/2} 
\end{eqnarray} 
to Eq. (\ref{gamma00}).
The result is the expressions of (\ref{bx2}) and
(\ref{gamma}). In doing so,
we have also made the use of Eq. (\ref{density99}).

\end{document}